\documentclass{appolb}

\usepackage{graphicx}
\usepackage{subfigure}
\usepackage{cite}

\newcommand{\lrd}{\raisebox{0.09em}{$\stackrel{\scriptstyle\leftharpoonup\hspace{-.7em}\rightharpoonup}{D} $}}

\newcommand{\ket}[1]{\left| #1 \right>}
\newcommand{\bra}[1]{\left< #1 \right|}

\begin{document}

\title{From chiral quark models to high-energy processes%
\thanks{Presented by WB at the {\em Cracow Epiphany Conference on hadron interactions at the dawn of the LHC}, 5-7 January 2009}%
\thanks{Dedicated to the memory of Jan Kwieci\'nski}%
}

\author{Wojciech Broniowski
\address{The H. Niewodnicza\'nski Institute of Nuclear Physics PAN, PL-31342 Krak\'ow\\
and Institute of Physics, Jan Kochanowski University, PL-25406~Kielce, Poland}
\and
Enrique Ruiz Arriola
\address{Departamento de F\'{\i}sica At\'omica, Molecular y Nuclear, Universidad de Granada, E-18071 Granada, Spain}
}
\maketitle
\begin{abstract}
We show the results of low-energy chiral quark models for soft matrix
elements involving pions and photons. Such soft elements, upon
convolution with the hard matrix elements, are relevant in various
high-energy processes. We focus on quantities related to the
generalized parton distributions of the pion: the parton distribution
functions, the parton distribution amplitudes, and the generalized
form factors.  Wherever possible, the model predictions are confronted
with the data or lattice simulations, where surprisingly good
agreement is achieved. The QCD evolution from the low quark model
scale up to the scale of the data is crucial for this agreement.
\end{abstract}

\PACS{12.38.Lg, 11.30, 12.38.-t}
 
\bigskip
\bigskip
 
Low-energy chiral quark models are designed to describe the the
dynamics of the pion and its interactions. Their treatment at the
large-$N_c$ level is particularly straightforward, as it requires a
rather simple task of computing the one-quark-loop diagrams with
insertions of pion or gauge-boson vertices. Despite numerous works
\cite{Davidson:1994uv,Dorokhov:1998up,Polyakov:1998td,Polyakov:1999gs,Dorokhov:2000gu,Anikin:2000th,Anikin:2000sb,RuizArriola:2001rr,Davidson:2001cc,RuizArriola:2002bp,RuizArriola:2002wr,Praszalowicz:2002ct,Tiburzi:2002kr,Tiburzi:2002tq,Theussl:2002xp,Broniowski:2003rp,Praszalowicz:2003pr,Bzdak:2003qe,Noguera:2005cc,Tiburzi:2005nj,Broniowski:2007fs,Courtoy:2007vy,Courtoy:2008af,Kotko:2008gy}
on the applications of low energy chiral quark models to high-energy
processes, their usefulness and predictive power is not well known to
the high-energy community. It has been widely believed that because
the non-perturbative structure of the pion is such a tremendously
difficult problem, it is better to parametrize the matrix elements
(for instance, the parton distribution function) by some suitable
form. Then this form is inserted as the initial condition to the QCD
evolution equations, which bring it to the relevant and experimentally
accessible scales. Clearly, that way one can only relate the
quantities at different scales and probe the evolution itself, but no
genuine non-perturbative dynamics present in the soft matrix elements
is investigated. In the case of the pion its low energy properties are
expected to be dominated by the spontaneous breakdown of the chiral
symmetry, a feature which is implemented in chiral quark models but is
largely ignored in many high energy studies.

In our approach two basic elements are crucial: the low-energy
dynamical quark model itself, as well as the QCD {em evolution},
moving the model predictions from the low quark model scale to higher
scales, where the experimental or lattice data are available.  Without
an operational definition of the low energy scale $Q_0$ and the QCD
evolution relating $Q_0$ to $Q$ one cannot compare the model
predictions to the data at the scale $Q$.  This talk is mainly based
on results published recently in
Refs.~\cite{Broniowski:2007si,Broniowski:2008hx}.

The theoretical framework is set by the Generalized Parton Distributions (GPDs) 
\cite{Ji:1998pc,Radyushkin:2000uy,Goeke:2001tz,Bakulev:2000eb,Diehl:2003ny,Ji:2004gf,Belitsky:2005qn,Feldmann:2007zz,Boffi:2007yc}.
For the case of the pion, the GPD for the non-singlet channel is defined as (we omit the gluon gauge link operators, absent in the 
light-cone gauge)
\begin{eqnarray}
&& \epsilon_{3ab}\,{\cal H}^{q,I=1}(x,\zeta,t) \!=\!\! \int
\frac{dz^-}{4\pi} e^{i x p^+ z^-}\!\!\!\! \left . \langle \pi^b (p') | \bar \psi (0)
\gamma^+ \psi (z) \, \tau_3 | \pi^a (p) \rangle \right |_{z^+=0,z^\perp=0}, \nonumber \\
\end{eqnarray}
while in the singlet case
\begin{eqnarray}
&& \delta_{ab}\,{\cal H}^{q,I=0}(x,\zeta,t)\!\!=\!\!\!\int\!\!
\frac{dz^-}{4\pi} e^{i x p^+ z^-}\!\!\!\!\left . \langle \pi^b (p') | \bar \psi (0)
\gamma^+  \psi (z) | \pi^a (p) \rangle \right |_{z^+=0,z^\perp=0} \nonumber \\
&& \delta_{ab}\,{\cal H}^{g}(x,\zeta,t)\!\!= \!\!\!\int\!\!
\frac{dz^-}{4\pi p^{+}} e^{i x p^+ z^-}\!\!\!\!\left . \langle \pi^b (p') | F^{+\mu} (0) 
F^{+}_{~\mu} (z) | \pi^a (p) \rangle \right |_{z^+=0,z^\perp=0}, \nonumber \\
\end{eqnarray} 
where the kinematics is set by $p'=p+q$, $p^2=m_\pi^2$, $q^2=-2p\cdot q=t$, and $\zeta = q^+/p^+$, which denotes the momentum transfer 
along the light cone. GPDs provide very rich information of the structure of hadrons, which may 
eventually come from such exclusive processes as 
$ep \to ep\gamma$, $\gamma p \to p l^+ l^-$, $ep \to ep l^+ l^-$,
or from the lattice calculations.

Formal properties of GPDs are most elegantly written in the symmetric notation: 
$\xi= \frac{\zeta}{2 - \zeta}$, $X = \frac{x - \zeta/2}{1 - \zeta/2}$, 
with $0 \le \xi \le 1$, $-1 \le X \le 1$, where one finds
\begin{eqnarray}
H^{I=0}(X,\xi,t)=-H^{I=0}(-X,\xi,t), \; H^{I=1}(X,\xi,t)=H^{I=1}(-X,\xi,t).
\end{eqnarray} 
For $X \ge 0$ one has 
\begin{eqnarray}
{\cal H}^{I=0,1}(X,0,0) = q^{S, NS}(X), \label{eq:pdf}
\end{eqnarray}
with $q(x)^i$ being the standard parton distribution functions (PDFs).
The following {\em sum rules} hold:
\begin{eqnarray}
\forall \xi : &&\int_{-1}^1 \!\!\!\!\! dX\, {H}^{I=1}(X,\xi,t) = 2 F_V(t), \nonumber \\
              &&\int_{-1}^1 \!\!\!\!\! dX\,X \, {H}^{I=0}(X,\xi,t) = 2\theta_2(t)-2\xi^2 \theta_1(t), \nonumber
\end{eqnarray}
where $F_V(t)$ is the {\em electromagnetic form factor}, while $\theta_1(t)$ and $\theta_2(t)$
are the {\em gravitational form factors} \cite{Donoghue:1991qv}, 
related, correspondingly, to the charge conservation and to the momentum sum rule in the deep inelastic scattering.

More generally, the {\em polynomiality} conditions~\cite{Ji:1998pc}, following from the Lorentz invariance, time reversal, and hermiticity, 
state that
\begin{eqnarray}
\int_{-1}^1 \!\!\!\!\! dX\,X^{2j} \, {H}^{I=1}(X,\xi,t) = 2\sum_{i=0}^j A_{2j+1,2i}(t) \xi^{2i}, \nonumber \\
\int_{-1}^1 \!\!\!\!\! dX\,X^{2j+1} \, {H}^{I=0}(X,\xi,t) = 2\sum_{i=0}^{j+1} A_{2j+2,2i}(t) \xi^{2i}, \nonumber
\end{eqnarray}
where $A$'s are the {\em generalized form factors (GFFs)}. 
These form factors can be also expressed as matrix elements of the form
\begin{eqnarray}
&& \bra{\pi^{+}(p')} \overline{u}(0)\, \gamma^{\{\mu}\, i \lrd\/^{\mu_1}
  i \lrd\/^{\mu_2} \dots i \lrd\/^{\mu_{n-1}\}} \,u(0) \ket{\pi^{+}(p)} = \nonumber \\
&& \;\;\;\;2 P^{\{\mu}P^{\mu_1} \dots P^{\mu_{n-1}\}} A_{n0}(t)
   + \nonumber \\ && 2\sum^n_{k=2,4,\dots}q^{\{\mu}
    q^{\mu_1} \dots q^{\mu_{k-1}}
    P^{\mu_{k}} \dots P^{\mu_{n-1}\}} \,2^{-k}A_{nk}(t),
\end{eqnarray}
where $P=p'+p$, hence the GPDs may be viewed as an infinite collection of GFFs.
The {\em positivity bound}~\cite{Pire:1998nw,Pobylitsa:2001nt} states that
\begin{eqnarray}
|H_q(X,\xi,t)| \le \sqrt{q(x_{\rm in}) q(x_{\rm out})}, \;\;\;\;\; \xi \le X \le 1, \label{positiv} 
\end{eqnarray}
where $x_{\rm in}=(x+\xi)/(1+\xi)$,  $x_{\rm out}=(x-\xi)/(1-\xi)$. 
Finally, a {\em low-energy theorem}~\cite{Polyakov:1998ze} 
\begin{eqnarray}
H_{I=1}(2z-1,1,0)=\phi(z) \label{eq:sumDA}
\end{eqnarray}
holds, relating the GPD to the  pion {\em distribution amplitude} (DA), denoted as $\phi$.

We stress that the above-listed relations and bounds impose severe constraints on the possible form of the GPDs.
All these formal requirements are naturally 
satisfied in our quark-model calculations \cite{Broniowski:2007si}.

\begin{figure}[tb]
\subfigure{\includegraphics[width=0.499\textwidth]{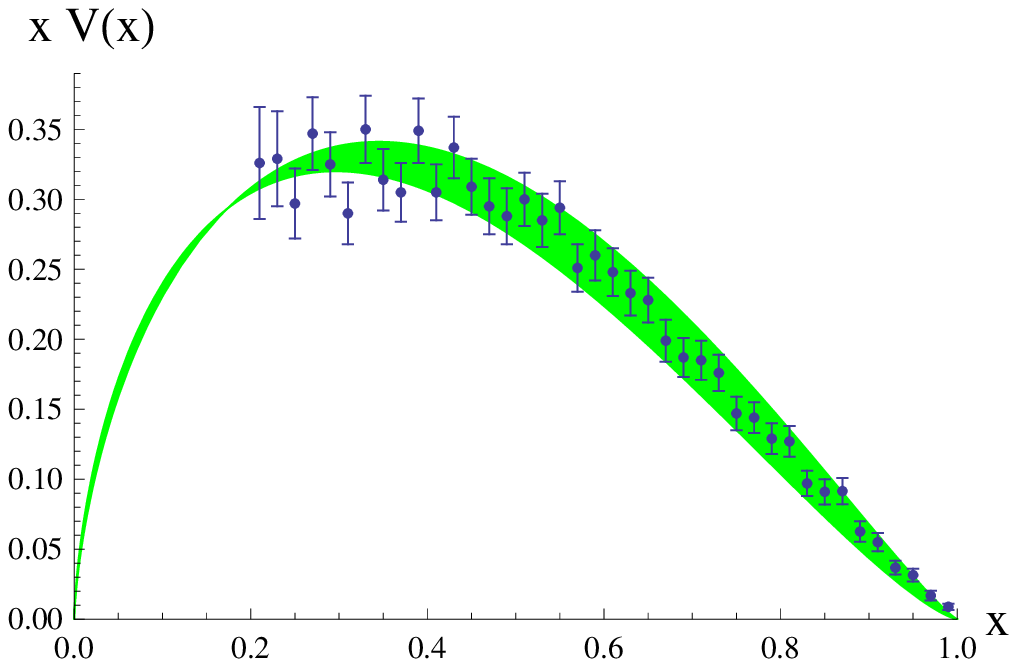}} \hfill
\subfigure{\includegraphics[width=0.499\textwidth]{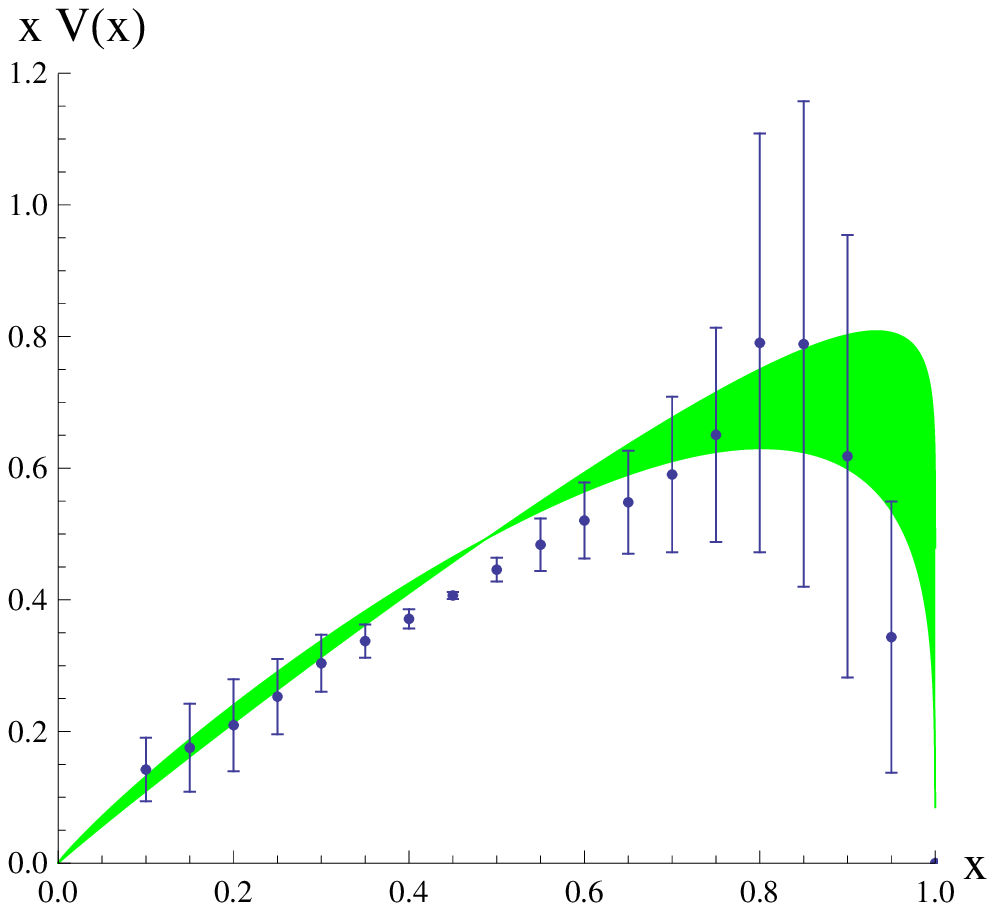}}
\caption{Left: model prediction for the valence
parton distribution in the pion evolved to the scale of $Q=4$~GeV (band). The
width of the band indicates the uncertainty in the initial scale
$Q_0$, Eq.~(\ref{eq:Q0}). The data points come from the
Drell-Yan  E615 experiment \cite{Conway:1989fs}. Right: same, evolved to the scale $Q=0.35~{\rm
GeV}$. The data come from the transverse lattice
calculations of Ref.~\cite{Dalley:2002nj} \label{fig:pdf}}
\end{figure}

With $\zeta=t=0$, the GPD becomes the usual PDF, cf. Eq.~(\ref{eq:pdf}). Long ago Davidson and one of us (ERA) found that
in the Nambu--Jona-Lasinio model~\cite{Davidson:1994uv} ${q(x)=1}$. This result pertains to the low-energy 
chiral-quark-model scale, which is a priori not known and has to be determined.
At this scale the gluons have been integrated out and effective quark degrees of freedom are the only 
degrees of freedom. Thus, all observables are saturated by the quark contribution, in particular, the momentum 
sum rule. From experiment, the momentum fraction carried by the valence quarks is~\cite{Sutton:1991ay,Gluck:1999xe}
\begin{eqnarray} 
\langle x \rangle_v = 0.47(2) {\rm ~~at~~} Q^2 = 4~{\rm GeV}^2. \label{eq:x}
\end{eqnarray} 
We evolve this value backward with the LO DGLAP equations down to the scale where quarks carry all the momentum, $\langle x \rangle_v = 1$. 
This procedure, assuming the matching of the quark model results and QCD at the quark model scale,  
yields the {\em quark model scale}
\begin{eqnarray}
{Q_0 = 313_{-10}^{+20} {\rm~MeV}}, \label{eq:Q0}
\end{eqnarray} 
where the range reflects the uncertainty in~(\ref{eq:x}).

We admit that such a low scale, where ${\alpha(Q^2_0)}/({2\pi})=0.34$, 
makes the evolution very fast for the scales close to the initial
value $Q_0$. This calls for improvement of the evolution. However, the NLO effects are 
small~\cite{Davidson:2001cc}. On the other hand, the removal of the Landau pole by means 
of the {\em analytic QCD} method \cite{Shirkov:1997wi,Stefanis:2000vd,Bakulev:2004cu,Bakulev:2005fp} results in a too slow evolution, 
incapable of reconciling the quark-model results with experimental data.  
Thus in our studies we employ the simplest LO DGLAP QCD evolution, baring in mind that it is good at intermediate 
values of $x$. 

The results of our procedure for the non-singlet PDF (valence quark distribution) of the pion are shown on the left side of 
Fig.~\ref{fig:pdf}. We have evolved the 
quark model result (\ref{eq:x}) from the scale (\ref{eq:Q0}) up to the scale $Q=4$~GeV corresponding to the E615 experiment \cite{Conway:1989fs}. 
We note a very reasonable agreement for all the range where the data are available. 
On the right side we show the same quantity evolved to the rather low scale of $Q=350$~MeV and confronted with the {\em transverse lattice} data 
\cite{Dalley:2002nj}. Again, the agreement is quite remarkable, showing the features of transverse-lattice calculation, designed to work at low-energy scales.

\begin{figure}[tb]
\subfigure{\includegraphics[angle=0,width=0.499\textwidth]{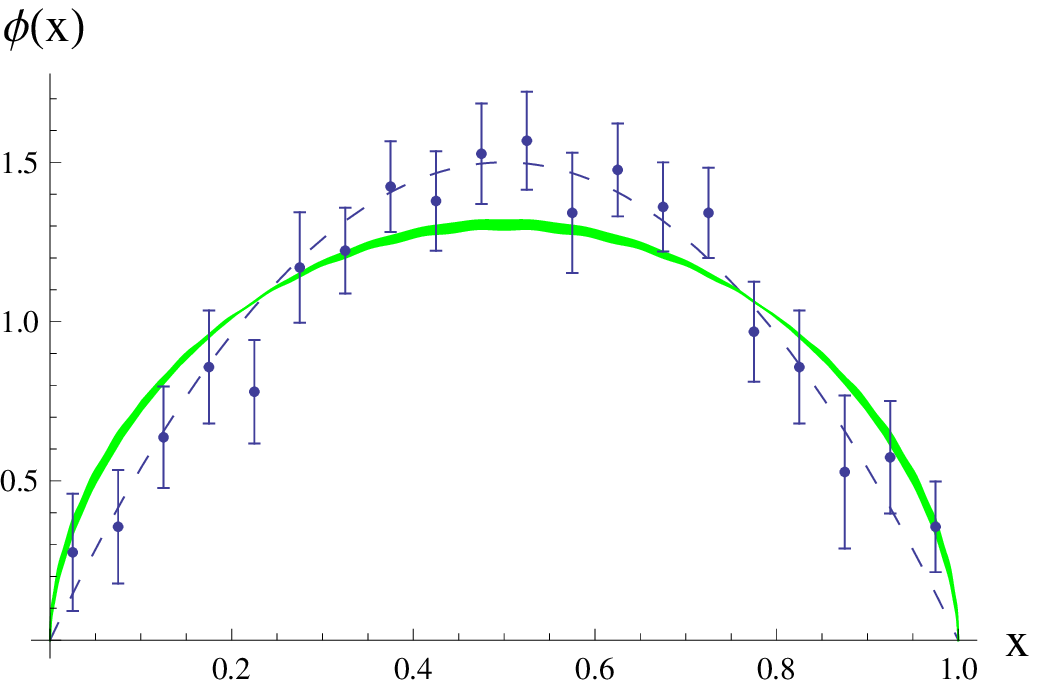}} \hfill
\subfigure{\includegraphics[angle=0,width=0.499\textwidth]{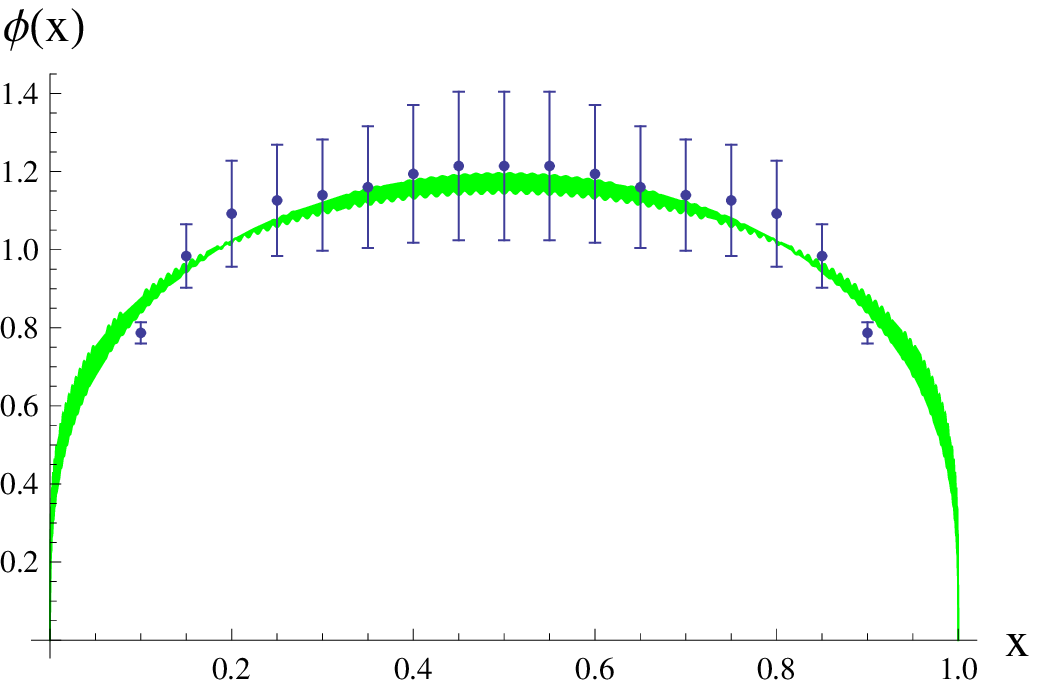}}
\caption{Left: model prediction for the pion distribution amplitude
evolved to the scale $Q=2~{\rm GeV}$ (band) and compared to the E791 di-jet
measurement~\cite{Aitala:2000hb}. The width of the band indicates the 
uncertainty in the initial scale $Q_0$, Eq.~(\ref{eq:Q0}). The dashed line shows
the asymptotic form $\phi(x,\infty)=6x(1-x)$. Right: same evolved to the scale 
$Q=0.5 {\rm GeV} $ compared to the transverse lattice data~\cite{Dalley:2002nj}.}
\label{fig:pda}
\end{figure}

Next, we look at the DA, related to the GPD via the sum rule
(\ref{eq:sumDA}). Here the evolution is carried out with the LO ERBL
equations. The results are displayed in Fig.~\ref{fig:pda}, again in
fair agreement with the data, especially for the lattice case on the
right-hand side.  Thus the GPDs for the special kinematic cases of
Eqs.~(\ref{eq:pdf},\ref{eq:sumDA}) are well reproduced in chiral quark
models {\em supplied with evolution}.

\begin{figure}[tb]
\subfigure{\includegraphics[width=.499\textwidth]{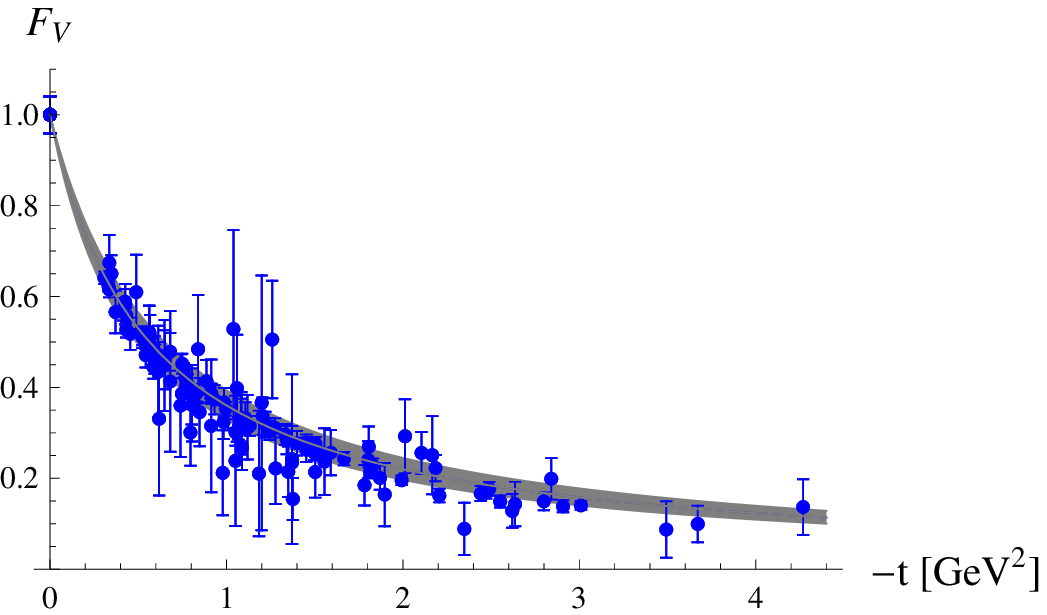}} \hfill
\subfigure{\includegraphics[width=.499\textwidth]{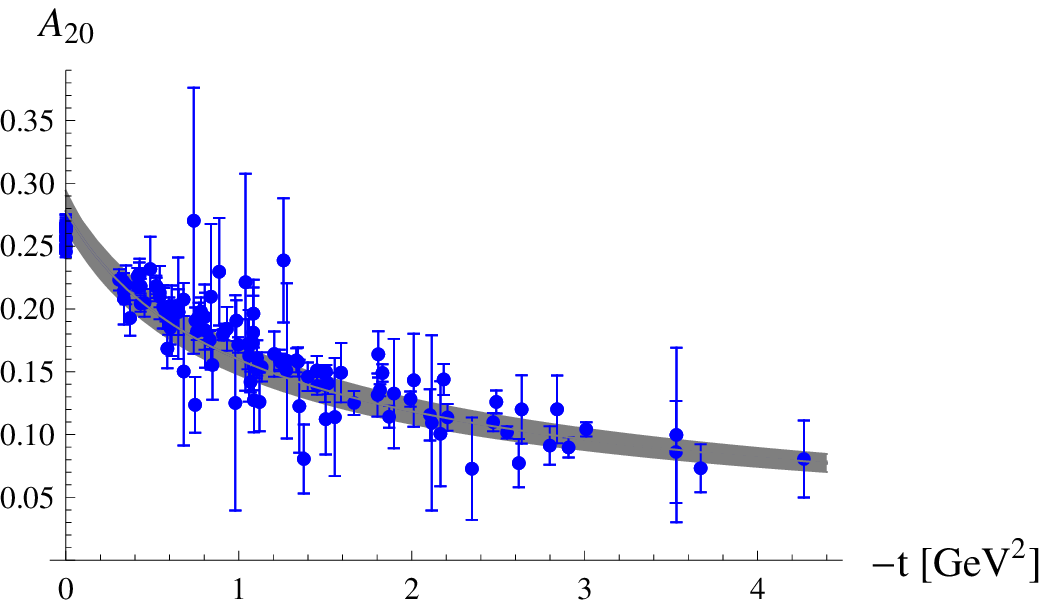}} 
\caption{The electromagnetic form factor (left) and the
  quark part of the gravitational form factor, $A_{20}$ (right) in the Spectral Quark Model compared to the lattice data from
  Ref.~\cite{Brommel:PhD}. The band around the model curve indicates
  the uncertainty in the quark momentum fraction and the
  $m_\rho$ parameter in the Spectral Quark Model. \label{fig:ff}}
\end{figure}

In Ref.~\cite{Broniowski:2007si} we provide expressions for the full
form of the GPDs in the NJL model and in the Spectral Quark Model
\cite{RuizArriola:2003bs}.  These expressions have a non-trivial
structure, in particular they {\em do not exhibit factorization} in
the $t$ and $x$ variables, while satisfying all formal requirements
listed above, including polynomiality.  Since there is no data for the
full kinematic range for the GPDs, we display the results for the
generalized form factors.  Interestingly, there is recent information
on these objects from the full QCD on the lattice
\cite{Brommel:PhD,Brommel:2005ee}.  The Spectral Quark Model vector
form factor and the quark part of the gravitational form factor of the
pion, $A_{20}$, are compared to these lattice data in
Fig.~\ref{fig:ff}.  We note a very good agreement. In the Spectral
Quark Model the expressions are particularly simple,
\begin{eqnarray}
F_V^{\rm SQM}(t)&=&\frac{m_\rho^2}{m_\rho^2-t}, \\\
\theta_1^{\rm SQM}(t)&=&\theta^{\rm SQM}_2(t)=\frac{m_\rho^2}{t} \log
\left ( \frac{m_\rho^2}{m_\rho^2-t} \right ). \nonumber 
\end{eqnarray}
We note the longer tail of the gravitational form factor in the momentum space, meaning a more compact distribution 
in the coordinate space. Explicitly, in the considered chiral quark models we find the simple relation
\begin{eqnarray}
2 \langle r^2 \rangle_\theta = \langle r^2 \rangle_V.
\label{eq:r2-V-theta}
\end{eqnarray} 

\begin{figure}[tb]
\begin{center}
\includegraphics[width=.73 \textwidth]{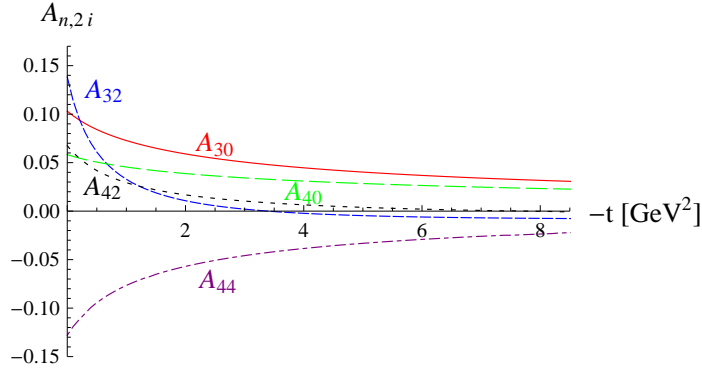}
\end{center}
\caption{(Predictions for the higher-order generalized form factors $A_{3,2i}$ and
  $A_{4,2i}$ of the pion from the Spectral Quark Model evolved to the scale $Q=2$~GeV. \label{fig:mom}}
\end{figure}

In Fig.~\ref{fig:mom} we show our predictions for the higher-order generalized form factors, evolved to the scale of 2~GeV.
In Ref.~\cite{Broniowski:evol} we have pointed out that these moments evolve in a very simple way with the LO DGLAP-ERBL evolution. One finds 
triangular structures, for instance in the non-singlet odd-$n$ case 
\begin{eqnarray}
&& A_{10}(t,Q)=L_1 A_{10}(t,Q_0), \nonumber \\
&& A_{32}(t,Q)=\frac{1}{5}(L_1-L_3)A_{10}(t,Q_0)+L_3 A_{32}(t,Q_0), \nonumber \\
&& A_{54}(t,Q)=\frac{1}{105}(9L_1-14L_3+5L_5)A_{10}(t,Q_0) \nonumber \\ && \;\;\;\;+\frac{2}{3}(L_3-L_5)A_{32}(t,Q_0)+L_5 A_{54}(t,Q_0), \nonumber \\
&& \dots \nonumber \\  
&& A_{30}(t,Q)=L_3 A_{30}(t,Q_0), \nonumber \\
&& A_{52}(t,Q)=\frac{2}{3}(L_3-L_5)A_{30}(t,Q_0)+L_5 A_{52}(t,Q_0), \nonumber \\
&& \dots \nonumber \\ 
&& A_{50}(t,Q)=L_5 A_{50}(t,Q_0),\nonumber \\
&& \dots  \label{eq:ffev}
\end{eqnarray}
where 
\begin{eqnarray}
L_n=\left ( \frac{\alpha(Q^2)}{\alpha(Q_0^2)}\right )^{\gamma_{n-1} /(2\beta_0)},
\end{eqnarray}
and $\gamma_i$ denotes anomalous dimensions, in the present case for
the vector operator.  Certainly, $L_1=1$, and the vector form factor,
corresponding to a conserved current, does not evolve. Similarly, the
gravitational form factors do not evolve. The higher-order form
factors, however, do change due to evolution.  Similar structures of
equations to (\ref{eq:ffev}) appear for the even $n$, as well as for
the singlet case \cite{Broniowski:evol}.  At the moment, there is no
data to compare to for the higher-order form factors, but our
predictions of Fig.~\ref{fig:mom} can be tested in future lattice
studies.

However, already at present there exists useful information at $t=0$.  Below 
we compare our model values of the higher-order form factors at
$t=0$ to the lattice data provided in Sec.~7 of
Ref.~\cite{Brommel:PhD}.  With the short-hand notation
$\langle x^n \rangle = A_{n+1,0}(0)$
one finds that at the lattice scale of $Q=2$~GeV
\begin{eqnarray}
&&\langle x \rangle = 0.271\pm 0.016, \\
&& \langle x^2 \rangle = 0.128\pm 0.018, \nonumber\\
&& \langle x^3 \rangle = 0.074\pm 0.027.\;\;\;(\rm lattice) \nonumber
\end{eqnarray} 
while in the chiral quark models we get, after the LO DGLAP
evolution to the lattice scale,
\begin{eqnarray}
&&\langle x \rangle = 0.28\pm 0.02, \\
&& \langle x^2 \rangle = 0.10\pm 0.02, \nonumber \\
&& \langle x^3 \rangle = 0.06\pm 0.01. \;\;\; (\rm chiral~quark~models) \nonumber
\end{eqnarray}
The model error bars come from the uncertainty of the scale $Q_0$ in
Eq.~(\ref{eq:Q0}). As we can see, the agreement falls within the error bars.

To briefly summarize the results presented in this talk, we state that
the chiral quark models supplied with the QCD evolution 
work well for a wide variety of quantities related to the GPDs of the pion. 
Thus they may provide valuable insight into the non-perturbative dynamics behind the 
soft matrix elements.
Further details can be found in Refs.~\cite{Broniowski:2007si,Broniowski:2008hx}.

\bigskip
\bigskip

Since at this meeting we are remembering Professor Jan Kwieci\'nski, I
wish to end my contribution by showing one of the last photos of our
dear teacher and friend sitting happily among his colleagues.  The
picture was taken on the occasion of handling to Jan the issue of Acta Physica Polonica B \cite{appb}
(placed at the corner of the table) dedicated to him in honor of his 65th birthday. The event took part 
in the Institute of Nuclear Physics in June 2003, just two and a half
months before his premature passing away. We all miss you, Jan!

\bigskip
\bigskip
Supported in part by the Polish Ministry of Science and Higher Education, grants N202~034~32/0918 and N202~249235, Spanish DGI and
FEDER funds with grant  FIS2008-01143/FIS, Junta de Andaluc{\'\i}a grant FQM225-05, and the EU Integrated Infrastructure Initiative
Hadron Physics Project, contract RII3-CT-2004-506078.


\begin{figure}[b]
\begin{center}
\includegraphics[width=.99\textwidth]{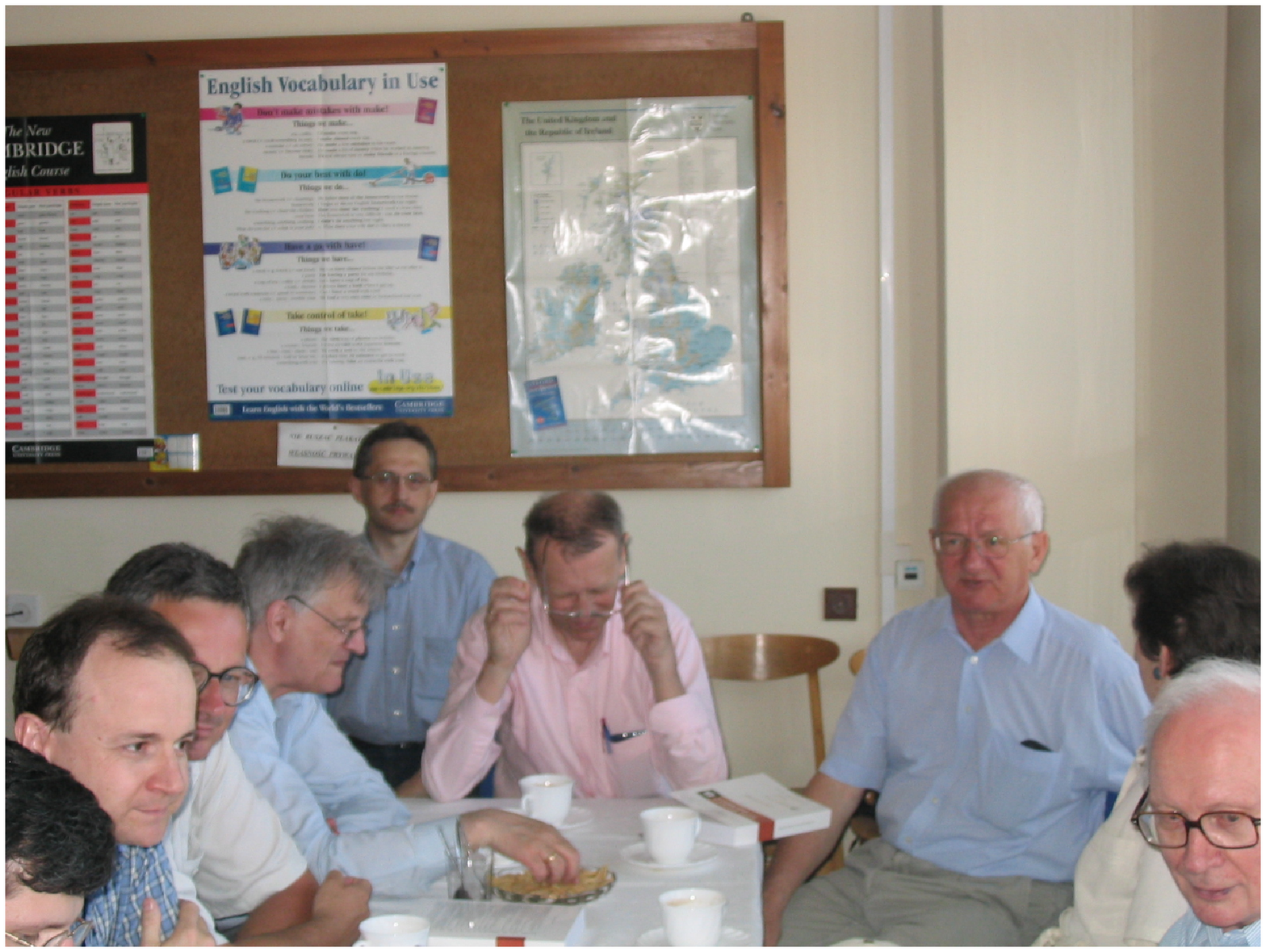}
\end{center}
\caption{12 June 2003, left to right: Mariusz Sadzikowski, Wojciech Florkowski, Micha\l{} Prasza\l{}owicz, Kacper Zalewski,
Krzysztof Golec-Biernat,  Andrzej Bia\l{}as, Jan Kwieci\'nski, Maria and Wies\l{}aw Czy\.z. \label{fig:jan}} 
\end{figure}

\end{document}